\documentstyle[epsfig]{aipproc}
\def\siml{\lower4pt \hbox{$\buildrel < \over \sim$}}
\def\simg{\lower4pt \hbox{$\buildrel > \over \sim$}}
\def\Mesz{M\'esz\'aros~}
\def\Pacz{Paczy\'nski~}

\def\nsns{NS-NS~}
\def\bhns{BH-NS~}

\def\eps{\epsilon}

\def\Fnu{F_{\nu}}

\def\Omj{\Omega_j}

\def\etal{{\it et~al.}}
\def\bec{\begin{center}}
\def\enc{\end{center}}
\def\beq{\begin{equation}}
\def\enq{\end{equation}}
\def\bea{\begin{eqnarray}}
\def\ena{\end{eqnarray}}

\newcommand{\boxsize}{0.89\textwidth}

\begin{document}
{\footnotesize
\noindent Review presented at the\hfill\\
{\it Cosmic Explosions}\hfill\\
10th October Astrophysics Conference\\
Maryland, Oct. 11-13 1999}
\bigskip

\title{ The Fireball Shock Model of \\ Gamma Ray Bursts }

\author{ P. \Mesz$^{1,2,3}$ } 
%\footnote{ E-mail address: nnp@astro.psu.edu} }

\address{$^1$Pennsylvania State University, 525 Davey, University Park, PA 16802 \\
$^2$California Institute of Technology, MS 105-24, Pasadena, CA 91125~~~~~~\\
$^3$E-mail address: nnp@astro.psu.edu} 

%\lefthead{LEFT head}
%\righthead{RIGHT head}

\maketitle

\begin{abstract}
Gamma-ray bursts are thought to be the outcome of a cataclysmic event leading to 
a relativistically expanding fireball, in which particles are accelerated at 
shocks and produce nonthermal radiation. We discuss the theoretical predictions of 
the fireball shock model and its general agreement with observations. Some of the 
recent work deals with the collimation of the outflow and its implications for the energetics, 
the production of prompt bright flashes at wavelenghts much longer than gamma-rays, 
the time structure of the afterglow, its dependence on the central engine or 
progenitor system behavior, and the role of the environment on the evolution of 
the afterglow.
\end{abstract}

\maketitle

\section{Introduction}

Gamma-ray bursts (GRB) have been studied in gamma-rays for over 25 years, but except for 
rare and fleeting X-ray detections, until a few years ago there existed no longer-lasting 
detections at softer wavelengths. However in early 1997 the Italian-Dutch satellite 
Beppo-SAX suceeded in providing accurate X-ray locations and images that allowed
their follow-up with large ground-based optical and radio telescopes.
The current interpretation of the gamma-ray and longer wavelength radiation is that 
the progenitor trigger produces an expanding relativistic fireball which can undergo 
both internal shocks leading to gamma-rays, and (as it decelerates on the external medium) 
an external blast wave and a reverse shock producing a broad-band spectrum lasting much longer.
%This fireball shock model has become the leading paradigm for the current understanding of GRB.

A strong confirmation of the generic fireball shock model came from the correct 
prediction \cite{mr97a}, in advance of the observations, of the quantitative nature 
of afterglows at longer wavelengths, in substantial agreement with the subsequent data 
\cite{vie97a,tav97,wax97a,rei97,wrm97}.  The measured $\gamma$-ray fluences imply a total 
energy of order $10^{54}(\Omega_\gamma /4\pi)$ ergs, where $\Delta \Omega_\gamma$ is the solid
angle into which the gamma-rays are beamed. Collimation may indeed be present, evidence 
having been recently reported for this \cite{kul99,fru99,cas99}. In any case,
such energies are possible \cite{mr97b} in the context of compact mergers
involving neutron star-neutron star (\nsns) or black hole-neutron star
(\bhns) binaries, or in hypernova/collapsar models involving a massive stellar
progenitor \cite{pac98,pop99}. In both cases, one is led to rely on MHD
extraction of the spin energy of a disrupted torus and/or a central 
BH to power a relativistic outflow.
 
\section{The Generic Fireball Shock Scenario}
\label{sec:fball}
 
Whatever the GRB trigger is, the ultimate result must unavoidably be an $e^\pm,\gamma$ 
fireball, which is initially optically thick. 
The initial dimensions must be of order $r_{min} \siml c t_{var} \sim 10^7$ cm, since 
variability timescales are $t_{var}\siml 10^{-3}$ s.  Most of the spectral energy 
is observed above 0.5 MeV, hence the $\gamma\gamma \to e^\pm$ mean free path is 
very short. Many bursts show spectra extending above 1 GeV, indicating
the presence of a mechanism which avoids degrading these via photon-photon interactions 
to energies below the threshold $m_e c^2=$ 0.511 MeV.
The inference is that the flow must be expanding with a very high Lorentz factor $\Gamma$, 
since then the relative angle at which the photons collide is less than $\Gamma^{-1}$ and 
the threshold for the pair production is diminished \cite{hb94}. 
However, the observed $\gamma$-ray spectrum is generally a broken power law, i.e., 
highly nonthermal. In addition, the expansion would lead to a conversion of internal 
into kinetic energy, so even after the fireball becomes optically thin, it would be 
radiatively inefficient, most of the energy being kinetic, rather than in photons.

The simplest way to achieve high efficiency and a nonthermal spectrum is
by reconverting the kinetic energy of the flow into random energy via shocks
after the flow has become optically thin \cite{rm92}.
Two different types of shocks may arise in this scenario. In the
first case (a) the expanding fireball runs into an external medium (the ISM, or a
pre-ejected stellar wind\cite{rm92,mr93a,ka94a,sapi95}. The second
possibility (b) is that \cite{rm94,px94}, even before external shocks occur,
internal shocks develop in the relativistic wind itself, faster portions of the
flow catching up with the slower portions.  This is a generic model, 
which is independent of the specific nature of the progenitor.

External shocks will occur in an impulsive outflow of total energy
$E_o$ in an external medium of average particle density $n_o$ at a radius
%\beq
$
r_{dec} \sim (3E_o/4\pi n_o m_p c^2 \eta^2)^{1/3} \sim 
 10^{17} E_{53}^{1/3} n_o^{-1/3} \eta_2^{-2/3} ~{\rm cm}~,
%\label{eq:rdec}
%\enq
$
and on a timescale $t_{dec} \sim  r_{dec}/(c\Gamma^2)$ $\sim 3\times 10^2  
E_{53}^{1/3} n_o^{-1/3}\eta_2^{-8/3}$ s,
where $\eta=\Gamma = 10^2\eta_2$ is the final bulk Lorentz factor of the ejecta.
Variability on timescales shorter than $t_{dec}$
may occur on the cooling timescale or on the dynamic timescale for
inhomogeneities in the external medium, but generally this is not ideal for
reproducing highly variable profiles\cite{sapi98} (see however \cite{dermit98}).
% (Sari \& Piran, 1997).
However, it can reproduce bursts with several peaks\cite{pm98a}
%(Panaitescu \& \Mesz, 1997)
and may therefore be applicable to the class of long, smooth bursts.
 
In a wind outflow \cite{pac90}, one assumes that a lab-frame luminosity $L_o$ and 
mass outflow $\dot M_o$ are injected at $r\sim r_l$ and continuously maintained over a time
$t_w$; here $\eta=L_o/ {\dot M_o c^2}$. In such wind model, internal shocks
will occur at a radius \cite{rm94}
$
%\beq
r_{dis} \sim  c t_{var} \eta^2 \sim 3\times 10^{14} t_{var} \eta_2^2 ~ {\rm cm},
%\label{eq:rdis}
%\enq
$
on a timescale $t_w \gg t_{var} \sim r_{dis}/(c\eta^2)$ s, 
where shells of different energies $\Delta \eta \sim \eta$ initially separated
by $c t_v $ (where $t_v \leq t_w$ is the timescale of typical variations in
the energy at $r_l$) catch up with each other.
In order for internal shocks to occur above
the wind photosphere $r_{ph} \sim {\dot M} \sigma_T /(4\pi m_p c \Gamma^2)$
$=1.2\times 10^{14} L_{53}\eta_2^{-3}$ cm, but also at radii greater than the
saturation radius (so the bulk of the energy does not appear in the
photospheric quasi-thermal component) one needs to have
$7.5\times 10^1 L_{51}^{1/5} t_{var}^{-1/5} \siml \eta
3\times 10^2 L_{53}^{1/4} t_{var}^{-1/4}$.
This type of models have the advantage\cite{rm94} that they allow an
arbitrarily complicated light curve, the shortest variation timescale $t_{var}
\simg 10^{-3}$ s being limited only by the dynamic timescale at $r_l$, where
the energy input may be expected to vary chaotically.
Such internal shocks have been shown explicitly  to
reproduce (and be required by) some of the more complicated
light curves\cite{sapi98,kps98,pm99int}.

A potentially valuable diagnostic tool for the central engine of GRB
is the power density spectrum (PDS). An analysis of BATSE light curves \cite{beloborodov+98}
%(Beloborodov \etal 1998) %(\cite{stern+98pds})
indicates that the logarithmic slope of the PDS between $10^{-2}$ and 2 Hz is approximately 
-5/3, and there is a cutoff of the average PDS above 2 Hz. 
\begin{figure}[ht]
\begin{center}
\begin{minipage}[t]{0.5\textwidth}
\epsfxsize=\boxsize
\epsfbox{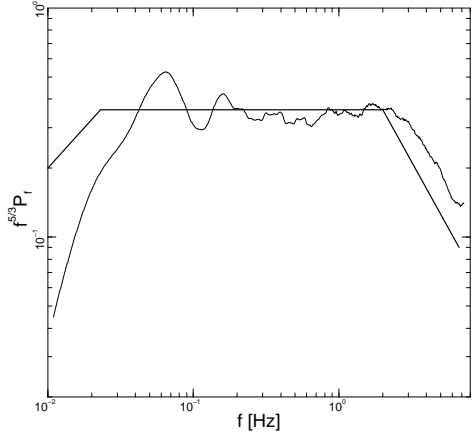}
\end{minipage}
\hspace{15mm}
\begin{minipage}[t]{0.35\textwidth}
\vspace*{-5cm}
\caption{\label{fig:pds} Average power density spectrum $P_f$ of simulated bursts from internal shocks, compared with the observed PDS (thick line), using a square-sine modulated Lorentz factor and a cosmological distribution satisfying the observed logN-logP (Spada, Panaitescu \& \Mesz 1999). }
\end{minipage}
\end{center}
\vspace*{-.5cm}
%\label{fig:pds}
\end{figure}
Using a simple kinematical model for the ejection and collision of relativistic shells
\cite{psm99,spm99} have calculated the
%Panaitescu, Spada \& \Mesz (1999) and Spada, Panaitescu \& \Mesz (1999) %\cite{panspamesz99}
light curves and PDS expected for a range of total burst energies and for a
total mass ejected and bulk Lorentz factor distribution compatible with the
internal shock scenario (Figure \ref{fig:pds}). The redshift distribution
also affects the PDS, and the observed logN-logP relation is used as a constraint.
For optically thin winds, a slope approaching -5/3 requires a non-random
Lorentz factor distribution, e.g. with an asymmetrical time modulation
so as to produce a larger number of collisions at low frequencies 
(see also \cite{beloborodov99}).
A cutoff at high frequencies ($\sim 2$ Hz) can be understood in
terms of shocks which increasingly occur below the scattering photosphere
of the outflow, or a deficit of energy in short pulses due to the modulation
of the Lorentz factors favoring shocks arising further out.

A significant fraction of bursts appear to have
low energy spectral slopes steeper than 1/3 in energy\cite{preece+98,crider+97}.
%(Preece \etal 1998, Crider \etal 1997, etc.). 
This has motivated consideration of a thermal or nonthermal\cite{liang+97,liang+99}
comptonization mechanism,
%While an astrophysical model where this mechanism would arise naturally has been  
%left largely unspecified, 
%Ghisellini \& Celotti 1999 
which can be put in the astrophysical context \cite{ghiscel99apjl} of internal shocks leading 
to self-regulated pair formation.
\begin{figure}[ht]
\begin{center}
%\begin{minipage}[t]{0.5\textwidth}
%\epsfxsize=\boxsize
\epsfbox{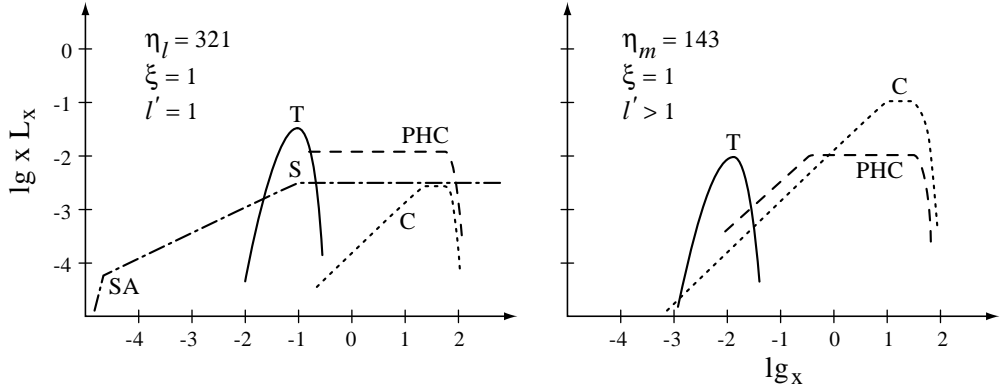}
%\end{minipage}
%\hspace{15mm}
%\begin{minipage}[t]{0.35\textwidth}
%\epsfxsize=\boxsize
%\vspace*{-5cm}
\caption{\label{fig:photsp} Luminosity per decade $xL_x$ vs. $x=h\nu/m_e c^2$ for two values of $\eta=L/{\dot M}c^2$ and marginal (left) or large (rigt) pair compactness.  T: thermal photosphere, PHC: photospheric comptonized component; S: shock synchrotron; C: shock pair dominated comptonized component (\Mesz \& Rees, 1999b).}
%\end{minipage}
%\vspace*{-.5cm}
\end{center}
%\label{fig:photsp}
\end{figure}
There is also evidence that the clustering of the break energy
of GRB spectra in the 50-500 keV range may not be due to observational selection
%(e.g. Preece \etal 1998; Brainerd \etal 1998; see however Dermer \etal 1999a).
\cite{preece+98,brainerd+98peak,dermer+99apjl}. % ,piranna96hunt}).
Models using Compton attenuation \cite{brainerd+98apj}  %(Brainerd \etal 1998) 
require reprocessing by an external medium whose column density adjusts itself to a 
few g cm$^{-2}$.  More recently a preferred break has been attributed to a blackbody
peak at the comoving pair recombination temperature in the fireball photosphere
\cite{eichlerlev99}.  %(Eichler \& Levinson 1999),
steep low energy spectral slope being due to the Rayleigh-Jeans part 
of the photosphere.  In order for such
photospheres to occur at the pair recombination temperature in the accelerating
regime requires an extremely low baryon load. For very large baryon loads, a
related explanation has been invoked \cite{tho94},
%Thompson (1994), 
considering scattering of photospheric photons off MHD turbulence in the coasting
portion of the outflow, which upscatters the adiabatically cooled photons
up to the observed break energy. These ideas have been synthesized \cite{mr99b}
%\Mesz \& Rees (1999b) %\cite{meszrees99b}
to produce a generic scenario (see Figure \ref{fig:photsp}) in which the presence
of a photospheric component as well as shocks subject to pair breakdown can produce
steep low energy spectra and preferred breaks. 

\section{Simple Standard Afterglow Model}
\label{sec:staaft}
 
The dynamics of GRB and their afterglows can be understood 
independently of any uncertainties about the progenitor systems, using
a generalization of the method used to model supernova remnants. The simplest
hypothesis is that the afterglow is due to a relativistic expanding blast wave,
which decelerates as time goes on \cite{mr97a}.
The complex time structure of some bursts suggests that the central trigger may
continue for up to 100 seconds, the $\gamma$-rays possibly being due to
internal shocks. However, at much later times all memory of the initial time
structure would be lost: essentially all that matters is how much energy and
momentum has been injected; the injection can be regarded as instantaneous in
the context of the afterglow.
As pointed in \cite{rm92}, the external shock
bolometric luminosity builds up as $L\propto t^2$ and decays as $L\propto 
t^{-(1+q)}$.  Beyond the deceleration radius
the bulk Lorentz factor decreases as a power law in radius, 
$
%\beq
\Gamma \propto r^{-g}\propto t^{-g/(1+2g)}~,~r\propto t^{1/(1+2g)},
%\label{eq:Gamma}
%\enq
$
with $g=(3,3/2)$ for the radiative or adiabatic regime (in which
$\rho r^3 \Gamma \sim$ constant or $\rho r^3 \Gamma^2 \sim$ constant).

The synchrotron peak frequency in the observer frame is $\nu_m \propto \Gamma B' \gamma^2$,
and both the comoving field $B'$ and electron Lorentz factor $\gamma$ are
expected to be proportional to $\Gamma$ \cite{mr93a}. As
$\Gamma$ decreases, so will $\nu_m$, and the radiation will move to longer
wavelengths. For the forward blast wave,
\cite{pacro93,ka94b} discussed the possibility of detecting at late times a radio or
optical afterglow of the GRB. A more detailed treatment of the fireball dynamics
indicates that approximately equal amounts of energy are radiated by the forward
blast wave, moving with $\sim\Gamma$ into the surrounding medium, and by a
reverse shock propagating with $\Gamma_r -1 \sim 1$ back into the ejecta
\cite{mr93a}. The electrons in the forward shock are hotter by a factor $\Gamma$ 
than in the reverse shock, producing two synchrotron peaks separated by $\Gamma^2$,
one peak being initially in the optical
(reverse) and the other in the $\gamma/X$ band (forward) \cite{mr93b,mrp94}. 
Detailed calculations and predictions of the time evolution of such a
forward and reverse shock afterglow model (\cite{mr97a}) preceded the
observations of the first afterglow GRB970228 (\cite{cos97,jvp97}), which
was detected in $\gamma$-rays, X-rays and several optical bands, and was
followed up for a number of months.
 
The simplest spherical afterglow model concentrates on the forward blast wave only.
For this, the flux at a given frequency and the synchrotron peak frequency decay at a 
rate \cite{mr97a,mr99}
\beq
\Fnu\propto  t^{[3-2g(1-2\beta)]/(1+2g)}~~,~~\nu_m\propto t^{-4g/(1+2g)},
\label{eq:Fnu}
\enq
where $g$ is the exponent of $\Gamma \propto r^{-g}$ and $\beta$ is
the photon spectral energy slope. The decay rate of the forward shock $\Fnu$
in equ.(\ref{eq:Fnu}) is typically slower than that of the reverse shock
\cite{mr97a}, and the reason why the "simplest" model was stripped down to
its forward shock component only is that, for the first two years 1997-1998,
afterglows were followed in more detail only after the several hours needed by
Beppo-SAX to acquire accurate positions, by which time both reverse external
shock and internal shock components are expected to have become unobservable.
This simple standard model has been remarkably successful at explaining
the gross features and light curves of GRB 970228, GRB 970508 (after 2 days; for
early rise, see \S \ref{sec:postaft})
e.g. \cite{wrm97,tav97,wax97a,rei97}. 
%(see Figure \ref{fig:0228_lightcurve}).
%Wijers, Rees \& \Mesz 1997, Tavani 1997, Waxman 1997, Reichart 1997).
%
%\begin{figure}[ht]
%\begin{center}
%\vspace*{-1.5cm}
%\begin{minipage}[t]{0.4\textwidth}
%\epsfxsize=\boxsize
%\epsfbox{sel970228.eps}
%\end{minipage}
%\hspace{15mm}
%\begin{minipage}[t]{0.4\textwidth}
%%\epsfxsize=\boxsize
%\vspace*{-4cm}
%\caption{The light curves of the afterglow of GRB 970228 at various wavelenghts, compared \cite{wrm97} to the simple blast wave model predictions of \cite{mr97a}.}
%\end{minipage}
%\vspace*{-1.5cm}
%\end{center}
%\label{fig:0228_lightcurve}
%\end{figure}
%

This simplest afterglow model has a three-segment power
law spectrum with two breaks. At low frequencies there is a steeply rising
synchrotron self-absorbed spectrum up to a self-absorption break $\nu_a$,
followed by a +1/3 energy index spectrum up to the synchrotron break $\nu_m$
corresponding to the minimum energy $\gamma_m$ of the power-law accelerated
electrons, and then a $-(p-1)/2$ energy spectrum above this break,
for electrons in the adiabatic regime (where $\gamma^{-p}$ is the electron
energy distribution above $\gamma_m$). A fourth segment and a third break is
expected at energies where the electron cooling time becomes short compared
to the expansion time, with a spectral slope $-p/2$ above that. With
this third ``cooling" break $\nu_b$, first calculated in \cite{mrw98} and
more explicitly detailed in \cite{spn98}, one has what has come to be called
the simple ``standard" model of GRB afterglows. One of the predictions of this
model \cite{mr97a} is that the relation between the temporal decay index $\alpha$,
for $g=3/2$ in $\Gamma\propto r^{-g}$, is related to the photon spectral energy
index $\beta$ through
$
%\beq
\Fnu \propto t^\alpha \nu^\beta~~,\hbox{with}~~\alpha=(3/2)\beta~.
%\label{eq:alphast}
%\enq
$
This relationship appears to be valid in many (although not all) cases, especially
after the first few days, and is compatible with an electron spectral index $p\sim
2.2-2.5$ which is typical of shock acceleration, e.g. \cite{wax97a,spn98,wiga98},
etc.  As the remnant expands the photon spectrum moves to lower frequencies, and
the flux in a given band decays as a power law in time, whose index can change
as breaks move through it.
For the simple standard model, snapshot overall spectra have been deduced
by extrapolating spectra at different wavebands and times using assumed
simple time dependences \cite{wax97b,wiga98}. These can be used to derive rough
fits for the different physical parameters of the burst and
environment, e.g. the total energy $E$, the magnetic and electron-proton
coupling parameters ${\eps}_B$ and ${\eps}_e$ and the external density $n_o$.
%(see Figure \ref{fig:0508_spec}).
%
%\begin{figure}[ht]
%\begin{center}
%\vspace*{-1.cm}
%\begin{minipage}[t]{0.4\textwidth}
%\epsfxsize=\boxsize
%\epsfbox{spec_0508.eps}
%\end{minipage}
%\hspace{35mm}
%\begin{minipage}[t]{0.3\textwidth}
%\epsfxsize=\boxsize
%\vspace*{-4cm}
%\caption{Snapshot spectrum of GRB 970508 at $t=12$ days and standard afterglow model fit \cite{wiga98}.}
%\end{minipage}
%\vspace*{-1.cm}
%\end{center}
%\label{fig:0508_spec}
%\end{figure}
%

\section{``Post-standard" Afterglow Models}
\label{sec:postaft}
 
%Despite the success of the simple standard model, realistically one
%could expect any of several fairly natural departures from this to occur. 
The most obvious departure from the simplest
standard model occurs if the external medium is inhomogeneous: for instance, for
$n_{ext} \propto r^{-d}$, the energy conservation condition is $\Gamma^2 r^{3-d}
\sim$ constant, which changes significantly the temporal decay rates \cite{mrw98}.
Such a power law dependence is expected if the external medium is a wind, say from
an evolved progenitor star as implied in the hypernova scenario (such winds are
generally used to fit supernova remnant models). 
Another departure from a simple impulsive injection approximation 
is obtained if the mass and energy injected during the
burst duration $t_w$ (say tens of seconds) obeys $M(>\Gamma) \propto
\Gamma^{-s}$, $E(>\Gamma)\propto \Gamma^{1-s}$, i.e. more energy emitted with
lower Lorentz factors at later times (but still shorter than the gamma-ray pulse
duration). This would drastically change the temporal decay rate and extend the
afterglow lifetime in the relativistic regime, providing a late ``energy refreshment"
to the blast wave on time scales comparable to the afterglow time scale
\cite{rm98}. These two cases lead to a decay rate
\beq
\Gamma \propto r^{-g} \propto \cases{
  r^{-(3-d)/2} & ~; $n_{ext}\propto r^{-d}$;\cr
  r^{-3/(1+s)} & ~; $E(>\Gamma)\propto \Gamma^{1-s}$.\cr }
\label{eq:Gammanonst}
\enq
Expressions for the temporal decay index $\alpha (\beta,s,d)$ in $\Fnu\propto
t^\alpha$ are given by \cite{mrw98,rm98}, which now depend also on $s$ and/or $d$.
The result is that the decay can be flatter (or steeper, depending on $s$ and $d$)
than the simple standard $\alpha= (3/2)\beta$.
A third non-standard effect, which is entirely natural, occurs when the energy
and/or the bulk Lorentz factor injected are some function of the angle. A simple case
is $E_o\propto \theta^{-j}$, $\Gamma_o\propto \theta^{-k}$ within a range of angles;
this leads to the outflow at different angles shocking at different radii and its
radiation arriving at the observed at different delayed times, and it has a marked
effect on the time dependence of the afterglow \cite{mrw98}, with $\alpha=\alpha
(\beta,j,k)$ flatter or steeper than the standard value, depending on $j,k$.
Thus in general, a temporal decay index which is a function of more than one
parameter
\beq
\Fnu\propto t^\alpha\nu^\beta~~,\hbox{with}~~\alpha=\alpha (\beta,d,s,j,k,\cdots )~,
\label{eq:alphanonst}
\enq
is not surprising; what is more remarkable is that, often, the simple
relation $\alpha=(3/2)\beta$ is sufficient to describe the overall
behavior at late times.
 
Evidence for departures from the simple standard model is provided by,
e.g., sharp rises or humps in the light curves followed by a renewed decay,
as in GRB 970508 (\cite{ped98,pir98a}). Detailed time-dependent model fits
\cite{pmr98} to the X-ray, optical and radio light curves of GRB 970228 and
GRB 970508 show that, in order to explain the humps, a {\it non-uniform} injection
%(Figure \ref{fig:injfit0508}) 
or an {\it anisotropic} outflow is required.
%
%\begin{figure}[ht]
%\begin{center}
%\vspace*{0.5cm}
%\begin{minipage}[t]{0.5\textwidth}
%\epsfxsize=\boxsize
%\epsfbox{curve0508.eps}
%\end{minipage}
%\hspace{15mm}
%\begin{minipage}[t]{0.4\textwidth}
%\vspace*{-3.5cm}
%\caption{\label{fig:injfit0508}
%Optical light-curve of GRB 970508, fitted with a non-uniform
%injection model (a similar fit can be obtained with an off-axis jet plus
%a weaker isotropic component) \cite{pmr98}.
%}
%\end{minipage}
%\end{center}
%\vspace*{-0.5cm}
%\end{figure}
These fits indicate that
the shock physics may be a function of the shock strength (e.g. the electron
index $p$, injection fraction $\zeta$ and/or $\epsilon_b,~\epsilon_e$ change
in time), and also indicate that dust absorption is needed to simultaneously
fit the X-ray and optical fluxes. The effects of beaming (outflow within a
limited range of solid angles) can be significant \cite{pmjet99}, but are coupled
with other effects, and a careful analysis is needed to disentangle them.
 
Prompt optical, X-ray and GeV flashes from reverse and forward shocks, as well as
from internal shocks, have been calculated in theoretical fireball shock models
for a number of years \cite{mr93b,mrp94,pm96,mr97a,sp99a}, as have been jets (e.g.
\cite{mr92,mlr93,mrp94}, and in more detail \cite{rho97,pmr98,pmjet99,rho99}).
However, observational evidence for these effects were largely lacking, until
the detection of a prompt (within 22 s) optical flash from GRB 990123 with
ROTSE by \cite{ake99}, together with X-ray, optical and radio follow-ups
\cite{kul99,gal99,fru99,and99,cas99,hjo99}. GRB 990123 is so far unique not only
for its prompt optical detection, but also by the fact that if it were emitting
isotropically, based on its redshift $z=1.6$ \cite{kul99,and99} its energy would
be the largest of any GRB so far, $4\times 10^{54}$ ergs. It is, however, also
the first (tentative) case in which there is evidence for jet-like emission
\cite{kul99,fru99,cas99}. An additional, uncommon feature is that a radio afterglow
appeared after only one day, only to disappear the next \cite{gal99,kul99}.
 
The prompt optical light curve of GRB 990123 decays initially as $\propto t^{-2.5}$
to $\propto t^{-1.6}$ \cite{ake99}, much steeper than the typical $\propto
t^{-1.1}$ of previous optical afterglows detected after several hours.
However, after about 10 minutes its decay rate moderates, and appears to
join smoothly onto a slower decay rate $\propto t^{-1.1}$ measured with
large telescopes \cite{gal99,kul99,fru99,cas99} after hours and days. The prompt
optical flash peaked at 9-th magnitude after 55 s \cite{ake99}, and in fact a
9-th magnitude prompt flash with a steeper decay rate had been predicted more than
two years ago \cite{mr97a}, from the synchrotron radiation of the reverse shock
in GRB afterglows at cosmological redshifts (see also {\cite{sp99a,mr93b,mrp94}). 
An origin of the optical prompt flash in internal shocks \cite{mr97a,mr99} cannot be ruled out, 
but is less likely since the optical light curve and the $\gamma$-rays appear not to correlate 
well \cite{sp99b,gal99}.  The subsequent slower decay agrees with predictions for the 
forward external shock \cite{mr97a,sp99b,mr99}.
 
The evidence for a jet is based on an apparent steepening of the light curve
after about three days \cite{kul99,fru99,cas99}. 
If real, this steepening is probably due to the transition between early
relativistic expansion, when the light-cone is narrower than the jet opening,
and the late expansion, when the light-cone has become wider than the jet,
leading to a drop in the effective flux \cite{rho97,kul99,mr99,rho99}. A rough
estimate leads to a jet opening angle of 3-5 degrees, which would reduce the total
energy requirements to about $4\times 10^{52}$ ergs. This is about two order of
magnitude less than the binding energy of a few solar rest masses, which, even
allowing for substantial inefficiencies, is compatible with currently favored
scenarios (e.g. \cite{pop99,mcfw99}) based on a stellar collapse or a compact
merger.
 
\section{Location and Environmental Effects}
\label{sec:env}
 
The location of the afterglow relative to the host galaxy center can
provide clues both for the nature of the progenitor and for the external
density encountered by the fireball. A hypernova model would be expected
to occur inside a galaxy in a high density environment $n_o > 10^3-10^5$ cm$^{-3}$.
Most of the detected and well identified afterglows are inside the projected image
of the host galaxy \cite{bloo98rome}, and some also show evidence for a dense
medium at least in front of the afterglow (\cite{ow98}).

In NS-NS mergers one would expect a BH plus debris torus system and
roughly the same total energy as in a hypernova model, but the mean distance
traveled from birth is of order several Kpc \cite{bsp99}, leading to a burst
presumably in a less dense environment. The fits of \cite{wiga98} to the
observational data on GRB 970508 and GRB 971214 in fact suggest external densities
in the range of $n_o=$ 0.04--0.4 cm$^{-3}$, which would be more typical of a
tenuous interstellar medium. These could be within the volume of the galaxy,
but for NS-NS on average one would expect as many GRB inside as outside. This is
based on an estimate of the mean NS-NS merger time of $10^8$ years.
%other estimated merger times (e.g. $10^7$ years, \cite{vdh92}) would give a burst much closer
%to the birth site. 
BH-NS mergers would occur in timescales $\sim 10^7$
years, and would be expected to give bursts inside the host galaxy
(\cite{bsp99}; see however \cite{fw98}). In at least one ``snapshot" standard
afterglow spectral fit for GRB 980329 \cite{reila98} the deduced external
density is $n_o\sim 10^3$ cm$^{-3}$. In some of the other detected afterglows
there is other evidence for a relatively dense gaseous environments, as
suggested, e.g. by evidence for dust \cite{rei98} in GRB970508,
the absence of an optical afterglow and presence of strong soft X-ray
absorption \cite{gro97,mur97} in GRB 970828, the lack an an optical
afterglow in the (radio-detected) afterglow (\cite{tay97}) of GRB980329, and
spectral fits to the low energy portion of the X-ray afterglow of several
bursts \cite{ow98}. 
One important caveat is that all afterglows found so far are based on Beppo-SAX
positions, which is sensitive only to long bursts $t_b \simg 20$ s \cite{hur98}.
This is significant, since it appears
likely that NS-NS mergers lead \cite{mcfw99} to short bursts with $t_b \siml 10$ s.
To make sure that a population of short GRB afterglows is not being
missed will probably need to await results from HETE \cite{hetepage} and from the
planned Swift \cite{swiftpage} mission, which is designed to accurately locate
300 GRB/yr.

\begin{figure}[ht]
\begin{center}
\vspace*{-0.5cm}
\begin{minipage}[t]{0.6\textwidth}
\epsfxsize=\boxsize
%\epsfbox{sh19_sp3.eps}
\epsfbox{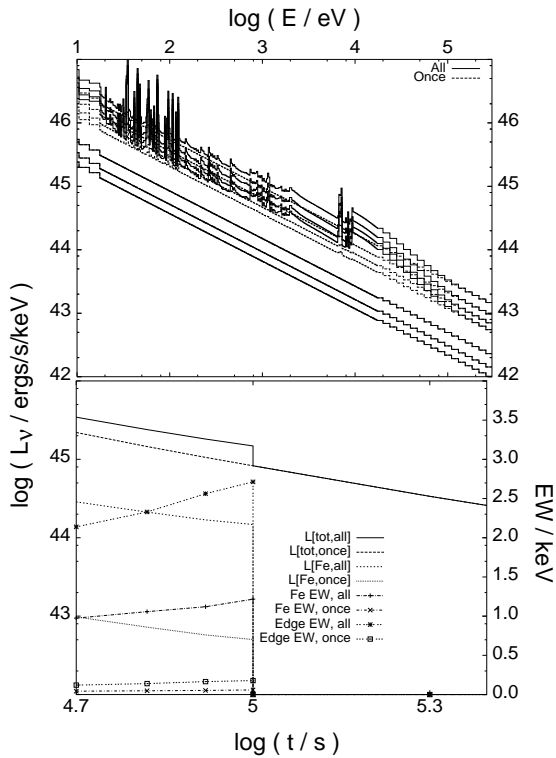}
\end{minipage}
\hspace{5mm}
\begin{minipage}[t]{0.29\textwidth}
\vspace*{-8cm}
\caption{\label{fig:shellsp} Spectrum (top) of a hypernova funnel model for various observer times, showing (bottom) the total and Fe light curves and equivalent widths (Weth, \Mesz, Kallman \& Rees 1999), for 
with $R=1.5\times 10^{16}$ cm, $n=10^{10}$ cm$^{-3}$, and Fe abundance $10^2$ times solar.}
\end{minipage}
\vspace*{-0.5cm}
\end{center}
%\label{fig:shellsp}
\end{figure}
The environment in which a GRB occurs may also lead to
specific spectral signatures from the external medium imprinted in the 
continuum, such as atomic edges and lines \cite{bkt97,pl98,mr98b}. These may be
used both to diagnose the chemical abundances and the ionization state (or local
separation from the burst), as well as serving as potential alternative redshift
indicators. (In addition, the outflowing ejecta itself may also contribute
blue-shifted edge and line features, especially if metal-rich blobs or filaments are
entrained in the flow from the disrupted progenitor debris \cite{mr98a}, which
could serve as diagnostic for the progenitor composition and outflow Lorentz factor).
To distinguish between progenitors, an interesting prediction
(\cite{mr98b}; see also \cite{ghi98,bot98}) is that the presence of a measurable
Fe  K-$\alpha$ X-ray {\it emission} line could be a diagnostic of a hypernova,
since in this case one may expect a massive envelope at a radius comparable to a
light-day where $\tau_T \siml 1$, capable of reprocessing the X-ray continuum by
recombination and fluorescence. Detailed radiative transfer calculations have
been performed to simulate the time-dependent X/UV line spectra of massive
progenitor (hypernova) remnants\cite{weth+99}, see Figure \ref{fig:shellsp}.
Two groups \cite{piro98b,yosh98} have in fact
recently reported the possible detection of Fe emission lines in GRB 970508 and
GRB 970828.

An interesting case is the apparent coincidence of GRB 980425 with the
unusual SN Ib/Ic 1998bw \cite{gal98_SN}, which may represent a new class of SN
\cite{iwa98,bloomSN98}. If true, this could imply that some or perhaps
all GRB could be associated with SN Ib/Ic \cite{wawe98}, differring only in
their viewing angles relative to a very narrow jet. Alternatively,
the GRB could be (e.g. \cite{wes98}) a new subclass of GRB with
lower energy $E_\gamma \sim 10^{48} (\Omj /4\pi )$ erg, only rarely observable,
while the great majority of the observed GRB would have the energies $E_\gamma
\sim 10^{54}(\Omj/4\pi)$ ergs as inferred from high redshift observations.
The difficulties are that it would require extreme collimations
by factors $10^{-3}-10^{-4}$, and the statistical association is so far not
significant \cite{kip98}. However, two more GRB light curves may have been
affected by an anomalous SNR (see, e.g. the review of \cite{whee99}).

\section{ Conclusions }
\label{sec:concl}
 
The fireball shock model of gamma-ray bursts has proved quite robust in
providing a consistent overall interpretation of the major features of these
objects at various frequencies and over timescales ranging from the short
initial burst to afterglows extending over many months. 
Significant progress has been made in understanding both the phenomenology and
the physics of these obejcts, which may the most widely studied type of black holes
sources.  There still remain a number of mysteries,
especially concerning the identity of their progenitors, the nature of the
triggering mechanism, the transport of the energy, the time scales involved,
and the nature and effects of beaming. However, the collective theoretical and 
observational understanding is vigorously advancing, and with dedicated new and 
planned observational missions under way, further significant progress  may be 
expected in the near future.
 
%\acknowledgements
\eject
I thank M.J. Rees, A. Panaitescu, M. Spada and C. Weth for stimulating 
collaborations, NASA NAG-5 2857, the Guggenheim Foundation, and the Division of Physics, 
Math \& Astronomy, Astronomy Visitor and Merle Kingsley funds at Caltech.

\end{document}